\documentclass[twocolumn]{aastex6}

\begin{document}

\title{The First High-Phase Observations of a KBO:\\New Horizons Imaging of (15810) 1994 JR$_1$ from the Kuiper Belt}

\author{
    Simon B. Porter\altaffilmark{1},
    John R. Spencer\altaffilmark{1},
    Susan Benecchi\altaffilmark{2},
    Anne J. Verbiscer\altaffilmark{3},
    Amanda M. Zangari\altaffilmark{1},
    H. A. Weaver\altaffilmark{4},
    Tod R. Lauer\altaffilmark{5},
    Alex H. Parker\altaffilmark{1},
    Marc W. Buie\altaffilmark{1},
    Andrew F. Cheng\altaffilmark{4},
    Leslie A. Young\altaffilmark{1},
    Cathy B. Olkin\altaffilmark{1},
    Kimberly Ennico\altaffilmark{6},
    S. Alan Stern\altaffilmark{1},
    and the New Horizons Science Team
}
\altaffiltext{1}{Southwest Research Institute, Boulder, CO, porter@boulder.swri.edu}
\altaffiltext{2}{Planetary Science Institute, Tucson, AZ}
\altaffiltext{3}{University of Virginia, Charlottesville, VA}
\altaffiltext{4}{Johns Hopkins Applied Physics Laboratory, Laurel, MD}
\altaffiltext{5}{National Optical Astronomy Observatory, Tucson, AZ}
\altaffiltext{6}{NASA Ames Research Center, Moffett Field, CA}

\begin{abstract}

    NASA's New Horizons spacecraft observed (15810) 1994 JR$_1$, a 3:2 resonant Kupier Belt Object (KBO),
    using the LOng Range Reconnaissance Imager (LORRI)
    on November 2, 2015 from a distance of 1.85 AU, and again on April 7, 2016 from a distance of 0.71 AU.
    These were the first close observations of any KBO other than Pluto.
    Combining ground-based and Hubble Space Telecope (HST) observations at small phase angles and the LORRI observations at higher phase angles, 
    we produced the first disk-integrated solar phase curve of a typical KBO from $\alpha$=0.6-58$^\circ$.
    Observations at these geometries, attainable only from a spacecraft in the outer Solar System, 
    constrain surface properties such as macroscopic roughness and the single particle phase function. 
    1994 JR$_1$ has a rough surface with a 37$\pm$5$^\circ$ mean topographic slope angle and has a relatively rapid rotation period of 5.47$\pm$0.33 hours.
    1994 JR$_1$ is currently 2.7 AU from Pluto; our astrometric points enable high-precision orbit determination and integrations which
    show that it comes this close to Pluto every 2.4 million years (10$^4$ heliocentric orbits), causing Pluto to perturb 1994 JR$_1$. 
    During the November spacecraft observation, the KBO was simultaneously observed using HST in two colors,
    confirming its very red spectral slope.
    These observations have laid the groundwork for numerous potential future distant KBO observations in the New Horizons-Kuiper Belt Extended Mission.

\end{abstract}


\section{Introduction} \label{sec:intro}

NASA's \textit{New Horizons} spacecraft launched on January 19, 2006, flew past Jupiter to obtain a gravity assist in February 2007, and passed within 13,700 km of Pluto on July 14, 2015 \citep{2015Sci...350.1815S}.
KBOs (Kuiper Belt Objects) form the larger and more primordial of the solar system's two circumstellar disks.
KBOs can only be observed from Earth within a few degrees of solar opposition, and a spacecraft is required to observe them at any higher solar phase angle.
The 2003 planetary science decadal survey ``New Frontiers in the Solar System'' \citep{Decadal2003} recommended that a ``Kuiper Belt-Pluto Explorer'' (which became \textit{New Horizons}) 
investigate several KBOs, and that the value of those investigations would be increased commensurately with the number of KBOs observed.

After the Pluto flyby, NASA approved a thruster burn sequence to redirect \textit{New Horizons} to the KBO 2014 MU$_{69}$, with the final burn on November 4, 2015.
This trajectory allows the spacecraft to reach 2014 MU$_{69}$ on January 1, 2019 \citep{2015LPI....46.1301P}.
On July 1, 2016 NASA approved the \textit{New Horizons} extended mission to 2014 MU$_{69}$\footnote{\url{
    https://www.nasa.gov/feature/new-horizons-receives-mission-extension-to-kuiper-belt-dawn-to-remain-at-ceres}}.
In addition to the 2014 MU$_{69}$ flyby, the planned extended mission also includes distant observations of about 20 KBOs at viewing geometries impossible from the Earth,
allowing close satellite and ring searches, measurement of surface properties, and precision orbit determination\footnote{\url{
    http://pluto.jhuapl.edu/News-Center/PI-Perspectives.php?page=piPerspective_04_14_2016}}.

(15810) 1994 JR$_1$ (hereafter JR1) is a Kuiper Belt Object in a 3:2 mean motion resonance with Neptune.
It was discovered in 1994 with the Issac Newton Telescope at the Roque de los Muchachos Observatory \citep{1995AJ....110.3082I},
making it the 13th KBO discovered (including Pluto-Charon) and the first resonant KBO after Pluto with a multi-year arc \citep{1996IAUS..172..153M}.
JR1 is currently 2.7 AU from Pluto and, like all 3:2 objects, is in a 1:1 resonance with Pluto.
\citet{2012MNRAS.427L..85D} proposed that this makes JR1 a quasi-satellite of Pluto, 
an object that spends an extended amount of time close to Pluto but is not gravitationally bound.

Here we present new observations of JR1 obtained with \textit{New Horizons} and HST on November 2, 2015 and by \textit{New Horizons} on April 7, 2016,
These observations enabled the first rotational period determination, an improved orbit fit, and a much extended solar phase curve.
These observations demonstrate the capability of \textit{New Horizons} to explore similarly distant Kuiper Belt objects in its proposed extended mission.

\section{Observations}

\subsection{New Horizons Observations}

The primary astronomical imager on New Horizons is the LOng Range Reconnaissance Imager (LORRI), 
which consists of a spacecraft-body-mounted 20.8 cm aperture Ritchey-Chretien telescope feeding a triplet field-flattening lens assembly, and on to an unfiltered back-illuminated CCD \citep{2008SSRv..140..189C}.
The CCD has 1024x1024 illuminated pixels, with an average plate scale of 4.963571$\pm$0.000038 $\mu$rad/pixel\footnote{\url{
    http://naif.jpl.nasa.gov/pub/naif/pds/data/nh-j_p_ss-spice-6-v1.0/nhsp_1000/data/ik/nh_lorri_v201.ti}}.
However, \textit{New Horizons} does not have reaction wheels and all pointing must be done with reaction control thrusters,
which can only actively maintain pointing to within 4 arcseconds.
Thus for exposures longer than 0.3 seconds, LORRI images are binned 4$\times$4 on the spacecraft to 256x256 pixels.
These 4$\times$4 images are much easier to downlink to Earth and have minimal loss of spatial precision for long exposures.
All the 4$\times$4 images presented here used an exposure time of 9.967 seconds.

LORRI first observed JR1 on November 2, 2015 in four sets of ten 4$\times$4 images, spaced one hour apart, and designated ``JR1{\_}LORRI{\_}306''.
At the time of the observation, \textit{New Horizons} was 1.85 AU from the KBO, and seeing it a solar phase angle of 26.7 degrees.
LORRI observed JR1 again on April 7, 2016, at a distance of 0.71 AU and solar phase angle of 58.5 degrees.
The observation sequence began with two ``JR1{\_}LORRI{\_}DEEP'' stacks of 24 4$\times$4 images, with the second starting 30 minutes after the first.
This was followed 30 minutes later by ``JR1{\_}LORRI{\_}LIGHTCURVE{\_}01'', a sequence of nine sets of three 4$\times$4 images, spaced 30 minutes apart.
Finally, ``JR1{\_}LORRI{\_}LIGHTCURVE{\_}02'' was another nine sets of three 4$\times$4 images, but spaced one hour apart.
During the April encounter, \textit{New Horizons} also obtained \linebreak ``JR1{\_}LORRI{\_}CLOSE'',
an experimental sequence of non-binned 1$\times$1 images with only the central 512x512 pixels returned.
Here we only present results of the 4$\times$4 observations, as JR1 did not appear to be visible in the 1$\times$1s.
All LORRI observations will be archived in NASA's Planetary Data System (PDS).

\begin{figure}
    \plotone{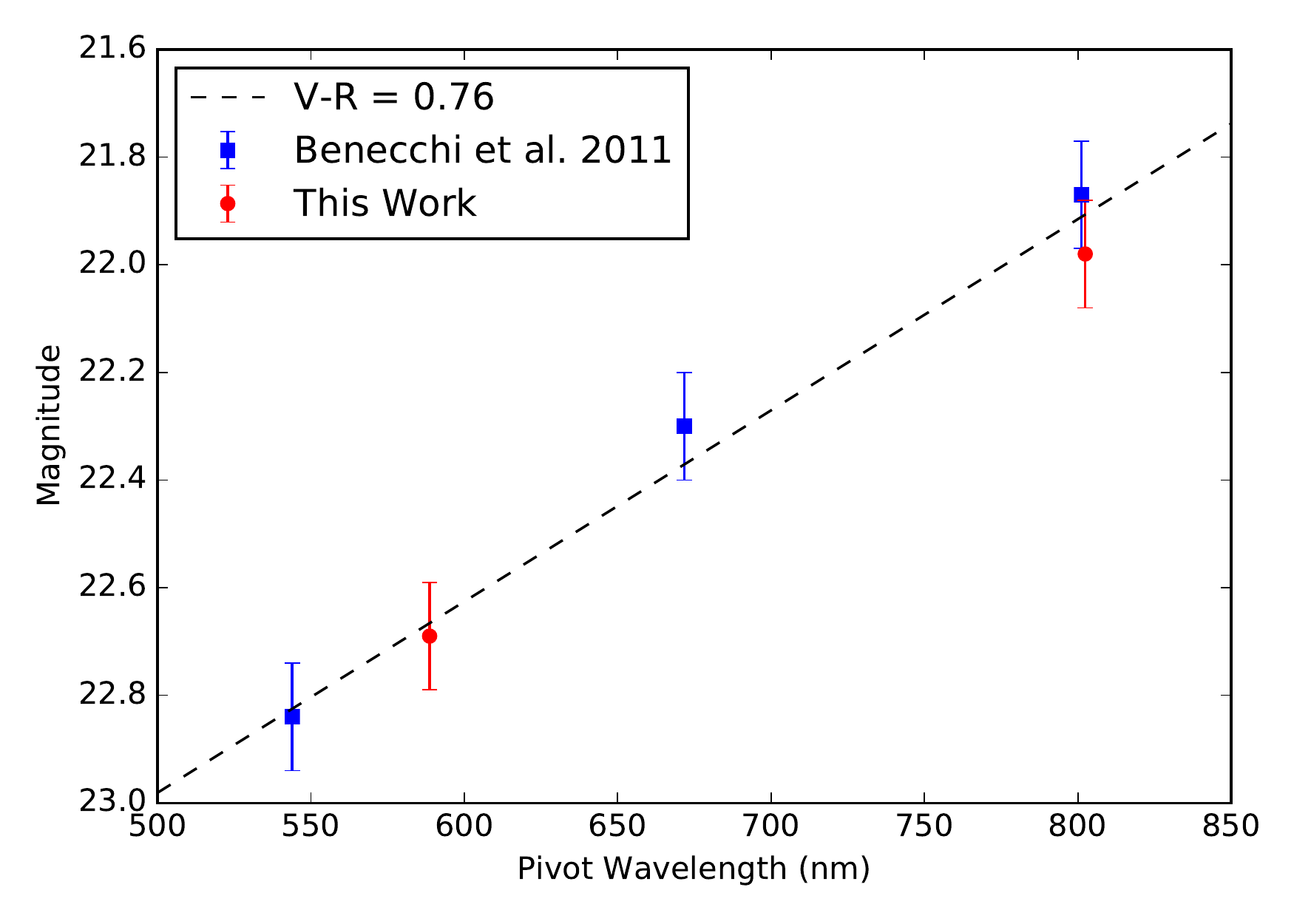}
    \caption{Measured colors for (15810) 1994 JR$_1$ with HST.
        Blue squares are observations by WFPC2 in F555W, F675W, and F814W published in \citet{2011Icar..213..693B}.
        Red circles are new WFC3 observations in F606W and F814W.
        The 0.1 magnitude error bars are an estimate of the lightcurve variation; actual measurement uncertainties are lower (see Table \ref{tab:lorri}).
        The dashed line corresponds to a spectral slope with V-R=0.76.
        These data have not been corrected for lightcurve variation.
    \label{fig:hst}}
\end{figure}

\floattable
\begin{deluxetable}{cclrcc}
    \tablecaption{LORRI Photometry of (15810) 1994 JR$_1$\label{tab:lorri}}
    \tablewidth{0pt}
    \tablehead{
        \colhead{Observation} & \colhead{Number of}    & \colhead{Right}     &                       &                       & \colhead{Solar} \\
        \colhead{Mid-Time}    & \colhead{Observations} & \colhead{Ascension} & \colhead{Declination} & \colhead{V Magnitude} & \colhead{Phase}
    }
    \startdata
2015-11-02 00:45:49 & 10 & 17$^h$14$^m$29.45$^s$ & -9:21:29.5 & $18.04\pm0.17$ & 26.715$^\circ$ \\
2015-11-02 01:45:49 & 10 & 17$^h$14$^m$28.93$^s$ & -9:21:21.6 & $18.17\pm0.20$ & 26.718$^\circ$ \\
2015-11-02 02:45:49 & 10 & 17$^h$14$^m$28.34$^s$ & -9:21:14.1 & $18.01\pm0.18$ & 26.721$^\circ$ \\
2015-11-02 03:45:49 & 10 & 17$^h$14$^m$27.80$^s$ & -9:21:06.9 & $18.35\pm0.19$ & 26.725$^\circ$ \\
2016-04-07 19:01:40 & 24 & 15$^h$41$^m$03.77$^s$ & 12:12:24.3 & $17.24\pm0.06$ & 58.210$^\circ$ \\
2016-04-07 19:31:40 & 24 & 15$^h$41$^m$01.88$^s$ & 12:12:49.1 & $17.41\pm0.07$ & 58.220$^\circ$ \\
2016-04-07 20:30:15 &  3 & 15$^h$40$^m$58.21$^s$ & 12:13:37.4 & $17.00\pm0.12$ & 58.240$^\circ$ \\
2016-04-07 21:00:15 &  3 & 15$^h$40$^m$56.31$^s$ & 12:14:02.5 & $16.98\pm0.15$ & 58.251$^\circ$ \\
2016-04-07 21:30:15 &  3 & 15$^h$40$^m$54.42$^s$ & 12:14:27.2 & $17.15\pm0.16$ & 58.261$^\circ$ \\
2016-04-07 22:00:15 &  3 & 15$^h$40$^m$52.53$^s$ & 12:14:51.9 & $17.50\pm0.20$ & 58.271$^\circ$ \\
2016-04-07 22:30:15 &  3 & 15$^h$40$^m$50.65$^s$ & 12:15:16.7 & $17.64\pm0.25$ & 58.282$^\circ$ \\
2016-04-07 23:00:15 &  3 & 15$^h$40$^m$48.73$^s$ & 12:15:41.1 & $17.10\pm0.15$ & 58.292$^\circ$ \\
2016-04-07 23:30:15 &  3 & 15$^h$40$^m$46.87$^s$ & 12:16:06.4 & $16.95\pm0.13$ & 58.302$^\circ$ \\
2016-04-08 00:00:15 &  3 & 15$^h$40$^m$44.95$^s$ & 12:16:31.3 & $17.08\pm0.13$ & 58.312$^\circ$ \\
2016-04-08 00:30:15 &  3 & 15$^h$40$^m$43.09$^s$ & 12:16:55.9 & $16.92\pm0.14$ & 58.323$^\circ$ \\
2016-04-08 01:30:15 &  3 & 15$^h$40$^m$39.29$^s$ & 12:17:45.7 & $17.29\pm0.16$ & 58.343$^\circ$ \\
2016-04-08 02:30:15 &  3 & 15$^h$40$^m$35.48$^s$ & 12:18:35.8 & $17.12\pm0.13$ & 58.364$^\circ$ \\
2016-04-08 03:30:15 &  3 & 15$^h$40$^m$31.72$^s$ & 12:19:25.6 & $17.39\pm0.17$ & 58.385$^\circ$ \\
2016-04-08 04:30:15 &  3 & 15$^h$40$^m$27.87$^s$ & 12:20:15.1 & $17.24\pm0.15$ & 58.405$^\circ$ \\
2016-04-08 05:30:15 &  3 & 15$^h$40$^m$24.10$^s$ & 12:21:05.3 & $17.09\pm0.14$ & 58.426$^\circ$ \\
2016-04-08 06:30:15 &  3 & 15$^h$40$^m$20.27$^s$ & 12:21:55.2 & $17.41\pm0.19$ & 58.447$^\circ$ \\
2016-04-08 07:30:15 &  3 & 15$^h$40$^m$16.50$^s$ & 12:22:45.2 & $16.86\pm0.13$ & 58.467$^\circ$ \\
2016-04-08 08:30:15 &  3 & 15$^h$40$^m$12.66$^s$ & 12:23:34.9 & $17.21\pm0.14$ & 58.488$^\circ$ \\
2016-04-08 09:30:15 &  3 & 15$^h$40$^m$08.83$^s$ & 12:24:25.2 & $17.59\pm0.24$ & 58.509$^\circ$ \\
    \enddata
    \tablecomments{Astrometric uncertainty is 0.6 arcseconds for all points.}
\end{deluxetable}

\begin{figure*}
    \plottwo{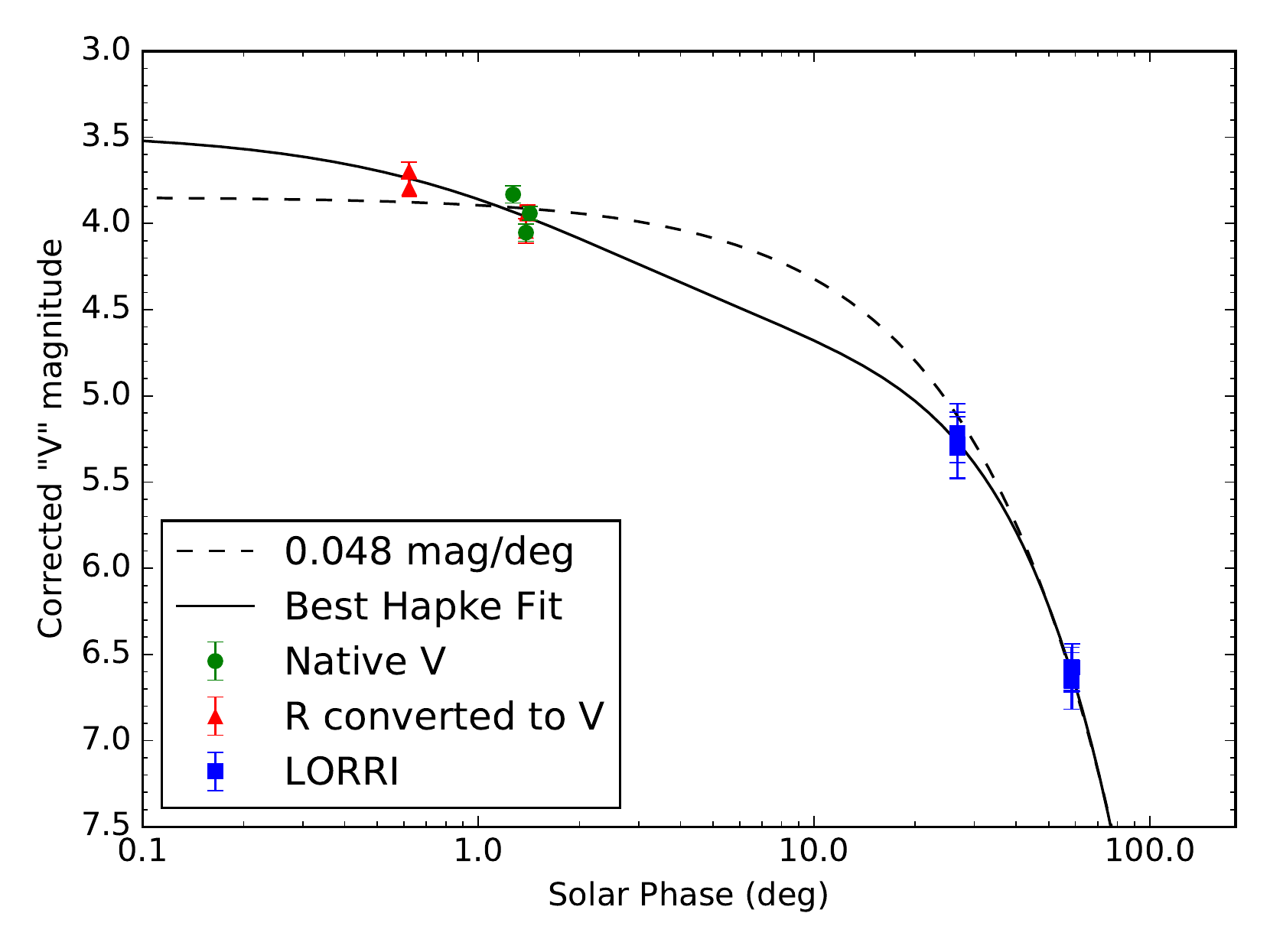}{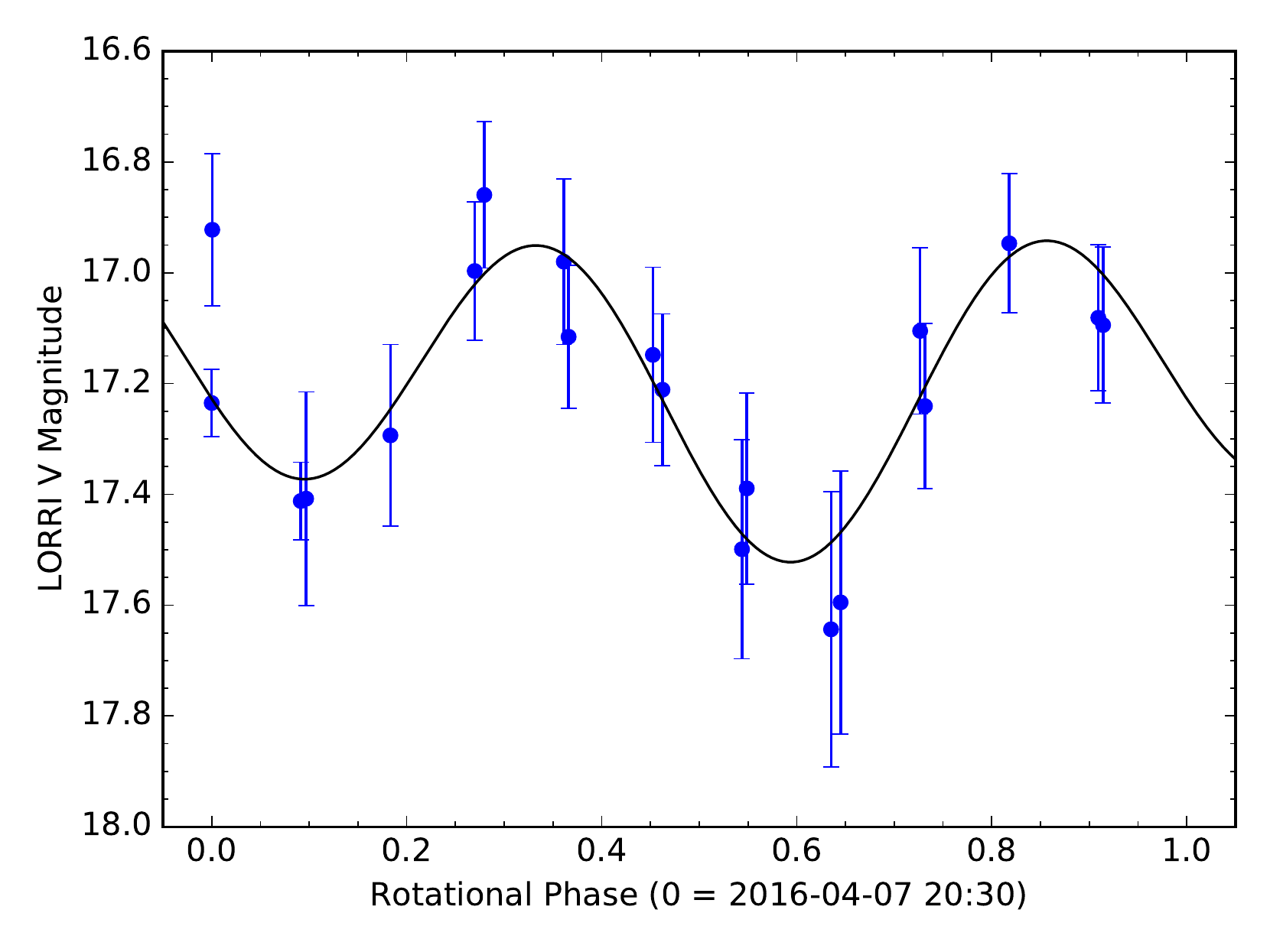}
    \caption{LORRI photometry of (15810) 1994 JR$_1$. 
        The left panel shows the best-fit phase curve and points from LORRI, \citet{1997MNRAS.290..186G}, \citet{2011Icar..213..693B}, and D. Tholen.
        R-band photometry was converted to V-band with V-R = 0.76 (see Figure \ref{fig:hst}).
        The right panel shows the April 2016 LORRI points phase-folded over a period of 5.47 hours, with median V magnitude 16.9 and peak-to-peak amplitude of 0.8 magnitudes.
        \label{fig:phot}}
\end{figure*}

Our first step was to build up a new delta-bias/hot pixel map to correct for bias drift \citep[see Fig. 10 in ][]{2008SSRv..140..189C}.
After subtracting this bias, we then extracted the point-like sources in the images and matched them to stars in the USNO UCAC4 catalog \citep{2013AJ....145...44Z}.
This allowed us to map the pointing of each image to a World Coordinate System (WCS) with a typical accuracy of 0.1 pixels (0.4 arcseconds).
Since JR1 was typically nearer the center of the field, we did not incorporate higher-order distortion coefficients.
Next, we reprojected each image into a common sidereal projection,
and then median combined the reprojected images to produce high signal-to-noise star stacks for both the November and April epochs.
This was especially important for the November epoch, as it had a much higher star density.
We next extracted windows from each original image, centered on the predicted location of the KBO using NAIF SPICE \citep{1996P&SS...44...65A},
scaled up by a factor of eight (i.e. twice the original pixel resolution), and rotated to the mean position angle for that set of observations.
We suppressed the stars by reprojecting the star stacks to the projection of each individual window and subtracting.
We median-combined each set of subtracted windows, which suppressed cosmic rays and other artifacts.
Between the star subtraction and the apparent sky motion of the KBO, we were able to suppress all but the brightest stars,
leaving just the KBO in the center of each stacked window.
Finally, we fit an elliptical 2-dimensional Gaussian to each stacked window, minimizing the $\chi^2$ residuals.
A non-circular form was needed because the spacecraft was not fully settled after each small slew for the lightcurve observations, causing the effective PSFs to be smeared in many cases.
We converted the pixel x/y location of the Gaussian to sky coordinates with the image's WCS
and measured the integrated DN/second by summing our model source.
We converted the DN/second to V magnitude with the calibrated 4$\times$4 zero point determined shortly after the Pluto flyby:
\begin{equation}
    V = 18.91 - 2.5*log_{10}(DN/second) + CC
\end{equation}
where CC is the color correction.
Of the objects with calibrated LORRI photomery, (5145) Pholus has the most similar spectral slope \citep[V-R=0.78;][]{1998Natur.392...49T}, 
and so we used its LORRI color correction of CC=0.213 magnitudes \citep{2008SSRv..140..189C}.
Note that this zero point is only valid for 4$\times$4 LORRI images.
The reduction pipeline and the orbit determination were repeated several times until the predicted location of the KBO matched the actual, meaning the KBO was centered in the windows.
This pipeline made extensive usage of the Astropy Python package \citep{2013A&A...558A..33A}.

\subsection{HST Observations}

In addition to the \textit{New Horizons} observations, the \textit{Hubble Space Telescope} (Program \#14092, S. Benecchi PI) observed JR1 on November 2, 2015,
starting just over a minute before the first LORRI observation (light-time corrected) and continuing for the following half hour. 
With this observation, the Wide Field Camera 3 (WFC3/UVIS) obtained five images, two with the F606W (wide V) filter and three with the F814W (wide I) filter.
The exposure time for the F606W images was 348 seconds, and 396 seconds for the F814W.
We performed the photometry for the HST observations using the same methodology as \citet{2011Icar..213..693B}.
This method matches the KBO's PSF with a synthetic model generated by Tiny Tim \citep{2011SPIE.8127E..0JK}, fitting for x and y positions, flux, and sky background.
We converted the x and y positions to sky coordinates with xytosky and used it in our orbit fits.
This was the first time a KBO (other than Pluto and Charon) was simultaneously observed from two extremely distant locations in the Solar System, 
and allowed us to very precisely constrain the present location of JR1.
The measured magnitude in F606W was $22.69\pm0.04$, and the magnitude in F814W was $21.98\pm0.02$.
These colors and complementary observations from HST/WFPC2 \citep{2011Icar..213..693B} are shown in Figure \ref{fig:hst}.
Differencing the November observations gives F606W-F814W = 0.71, while Figure \ref{fig:hst} shows an approximate V-R slope of $0.76\pm0.09$.
In either case, JR1 is a very red object.
The mean V-R (Johnson V, Cousins R) for the 12 3:2 resonant objects in \citet{2011Icar..213..693B} is 0.62, and JR1 is redder than 2/3 of those objects,
while Pluto's mean V-R is 0.56-0.60 \citep{2015ApJ...804L...6B}.
\citet{2015Icar..252..311D} showed that the surfaces of similarly red KBOs and Centaurs are composed of irradiated CH$_4$, CO, CO$_2$, and H$_2$O ices,
producing a coating of longer-chain hydrocarbons.

\begin{table}[t]
    \renewcommand{\thetable}{\arabic{table}}
    \centering
    \caption{Orbit and Physical Properties of 1994 JR$_1$} \label{tab:prop}
    \begin{tabular}{cr>{\centering\arraybackslash}p{0cm}l}
        \tablewidth{0pt}
        \hline
        \hline
        \multicolumn{4}{l}{Mean and Standard Deviation of $\pm$10 Myr Barycentric Orbit} \\
        \hline
        Semimajor Axis &  39.41 &$\pm$& 0.11 AU    \\
        Eccentricity &  0.1244 &$\pm$& 0.0058  \\
        \multicolumn{1}{>{\centering\arraybackslash}m{3cm}}{Inclination to Invariant Plane\tablenotemark{a}}&  3.48 &$\pm$& 0.60$^\circ$     \\
        \hline
        \multicolumn{4}{l}{Physical Properties}\\
        \hline
        V-R Color & 0.76 &$\pm$& 0.09 \\
        Rotational Period & 5.47 &$\pm$& 0.33 hours\\
        Assumed Diameter & \multicolumn{2}{c}{250 km} \\
        Geometric Albedo & \multicolumn{2}{c}{0.04} \\
        Phase Integral $q$ & \multicolumn{2}{c}{0.22} \\
        Phase Coefficient &  0.0475 &$\pm$& 0.0007 mag/deg \\
        \multicolumn{1}{>{\centering\arraybackslash}m{3cm}}{Single-Scattering Albedo $w$} & 0.025 &$\pm$& 0.005 \\
        \multicolumn{1}{>{\centering\arraybackslash}m{3.5cm}}{Surface Roughness $\bar{\theta}$} & 37 &$\pm$& 5$^\circ$ \\
        Henyey-Greenstein $\xi$ & -0.38 &$\pm$& 0.05 \\
        \hline
        \multicolumn{1}{l}{Opposition Surge} & Width & \multicolumn{2}{c}{Amplitude} \\
        \hline
        Shadow Hiding & 0.029 & \multicolumn{2}{c}{0.89} \\
        Coherent Backscatter & 0.026 & \multicolumn{2}{c}{0.73} \\
        \hline
    \end{tabular}
    \tablenotetext{a}{Assumed Invariant Pole: RA~23$^h$00$^m$31.9$^s$, Dec~-3:51:09.4}
\end{table}

\section{Photometry}

After its discovery in 1994, JR1 was observed a number of times in the 1990s.
The best-calibrated photometry of this period was obtained in both V and R in 1995 and published in \citet{1997MNRAS.290..186G}.
\citet{2011Icar..213..693B} obtained additional color photometry with HST/WFPC2 in 2001 and HST/ACS in 2002.
After a 13-year gap, JR1 was recovered by D. Tholen (MPS 610223) in June 2015 with the University of Hawaii 2.2-meter telescope.
This was followed by the colors presented above obtained in November 2015 with HST/WFC3.
As described above, \textit{New Horizons} obtained four photometric points on November 2, 2015 at a cadence of one hour.
The spacecraft obtained twenty additional photometric points April 7, 2016, the first eleven at a cadence of 30 minutes, the last nine at a cadence of one hour.
These points are shown in Table \ref{tab:lorri}.

\subsection{Solar Phase Curve}

JR1 was an ideal first post-Pluto target for \textit{New Horizons}, given both its proximity to Pluto and that it was bright enough to be seen by the spacecraft at phase angles of 27$^\circ$ and 58$^\circ$.
The ground-based observations of \citet{1997MNRAS.290..186G} and HST observations in \citet{2011Icar..213..693B} and this work are all in the solar phase angle range $\alpha=$1-2$^\circ$.
The lowest solar phase photometry we considered were D. Tholen's June 2015 observations at $\alpha = 0.6^\circ$.
Future ground-based photometry closer to opposition are necessary to constrain the opposition surge fully.

Using the ground-based, HST, and LORRI observations, we generate JR1's disk-integrated solar phase curve (Figure \ref{fig:phot}) and fit the \citet{hapke2012theory} photometric model. 
The LORRI observations acquired in November 2015 and April 2016 have been corrected for rotational lightcurve variation to the mean of the double-peaked lightcurve shown in Figure \ref{fig:phot}.
The best-fit photometric parameters are summarized in Table \ref{tab:prop}.
While we do not have a constraint on the size of JR1 (as no occultations or thermal measurements have been reported),
we assume a mean diameter of 250 km, which gives a geometric albedo of 0.04, typical for a resonant KBO of this size \citep{2008ssbn.book..161S}.
The phase integral is 0.22, estimated using Russell's Rule \citep{1988Icar...73..324V}, giving a spherical, or Bond albedo of 0.01.
A linear fit to these points gives a phase coefficient of 0.0475$\pm$0.0007 magnitudes/degree.
This is smaller than most of the larger KBOs listed in \citet{2008ssbn.book..115B}, but similar to smaller Centaur (2060) Chiron.
The \citet{hapke2012theory} parameters are then single scattering albedo $w$=0.025, surface roughness (mean topographic slope angle) $\bar{\theta}$=37$^\circ \pm 5^\circ$.
and the Henyey-Greenstein \citep{1941ApJ....93...70H} single particle scattering asymmetry parameter (mean cosine of the scattering angle) $\xi$=-0.38.
This is a moderately-backscattering $\xi$, typical for icy objects in the outer solar system \citep{2013ASSL..356...47V}.
The opposition surge is characterized by angular widths and amplitudes contributed from both shadow hiding and coherent backscatter.
The angular width of the shadow hiding opposition surge is 0.029 and the amplitude is 0.89; the angular width of the opposition surge
due to coherent backscatter is 0.026 and the amplitude is 0.73.
The LORRI observations at the larger phase angles place constraints on the surface roughness and the single particle phase function.
This surface roughness is even higher than Saturn's dark, similarly-sized moon Phoebe
\citep[$\bar{\theta}$=32$^\circ$;][]{1999Icar..138..249S}.
This indicates that like Phoebe \citep[which is thought to have originated in the Kuiper Belt,][]{2005Natur.435...69J}, 
JR1 may have a very heavily cratered surface.

\subsection{Rotational Period}

Because of its infrequent cadence, the ground-based photometry of JR1 never showed a discernible periodicity.
The two lightcurve observations by \textit{New Horizons} were designed to measure any high-frequency variation by sampling at 30 minute and one hour cadences.
The 30-minute observations were downlinked first, and immediately displayed a periodic variation of around five hours (double-peaked).
When the rest of the April data were down, we fit an initial period using just the April photometry, as it was all at roughly the same distance and solar phase angle.
We iteratively phased the lightcurve between 3 and 14 hours, with steps of 2.88 seconds.
For each potential period, we derived a minimized $\chi^2$ two-term Fourier series, and selected the period with the lowest $\chi^2$ value.
Our best fit to just the April photometry has a period of $5.47\pm0.33$ hours and amplitude of 0.8 magnitudes,
and is shown in Figure \ref{fig:phot}.
This period is consistent with the variation in brightness seen in the four November spacecraft observations, 
though the period is uncertain enough to allow the two epochs to be combined.
A period of 5.47 hours is faster than the average period of 7.5 hours for 29 KBOs and Centaurs in \citet{2010A&A...522A..93T},
but well within their range of 3.56 to 12 hours for sub-dwarf planet sized objects.
Future ground-based high-cadence photometry (sampling several times over the period) would be very useful to reduce the uncertainty in the period 
and allow preliminary estimates of JR1's rotational pole and shape.

\section{Orbit}

We combined our astrometry of JR1 from \textit{New Horizons} with the previous Earth-based astrometry and our November 2015 HST observations
to produce an astrometric set that spanned almost 22 years and observer distances of 43 to 0.7 AU.
For the observer location from \textit{New Horizons}, we used the spacecraft trajectory solution OD124, with includes the burn to 2014 MU$_{69}$ in November 2015.
For Earth and HST observer locations, we used JPL DE433 ephemeris and HST kernels, respectively.

We then used emcee \citep{2013PASP..125..306F} to create a discretely-sampled probability distribution cloud of 100,000 orbital solutions by modeling all the available astrometric data 
(using appropriate uncertainties) with fully-perturbed ephemerides.
To do this, we propagated the orbits with a high-precision N-Body integrator
\footnote{\url{https://github.com/ascendingnode/PyNBody}} based on the 12/13th order Runge-Kutta-Nystrom intergrator of \citet{Brankin_1989}.
We used planetary masses and initial states from JPL DE433, substituting the more accurate Pluto mass from \citet{2015Icar..246..317B}.
The uncertainty in our orbit solution is very low, due both to the unique observational location of \textit{New Horizons} and the exceptionally long orbital arc for a KBO.
The major advantage of observing JR1 from the Kuiper Belt was in restricting the object's distance from the Sun, 35.50 AU at the time of these observations.
We were able to reduce the 1-$\sigma$ radial uncertainty in the position of JR1 from $10^5$ km with just Earth-based observations to less than $10^3$ km with the \textit{New Horizons} points.
In addition, because of JR1's early discovery and thus very long data arc for such a small KBO, we can also place tight constraints on its motion.
This then allowed us to run high-precision orbital integrations to test its past and future interactions with Pluto.

\begin{figure}
    \plotone{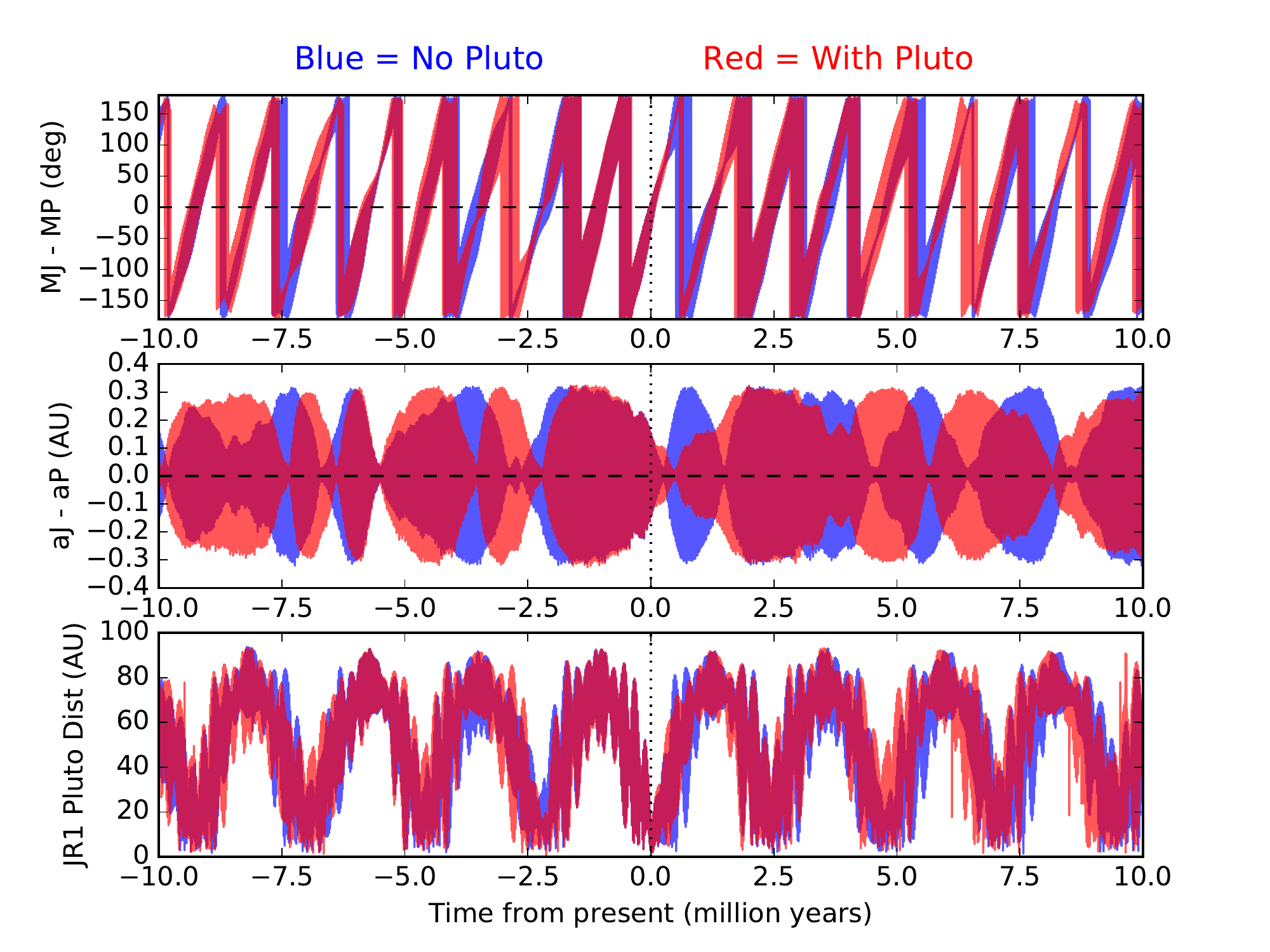}
    \caption{Relative orbital evolution for 1994 JR$_1$ and Pluto.
        The top panel show difference in their mean longitudes, the middle shows the difference in their semimajor axes,
        and the bottom shows the physical distance between the objects.
        When Neptune perturbations bring the two objects close to each other every 2.4 million years,
        Pluto does perturb JR1's orbit.
    \label{fig:evo}}
\end{figure}

\subsection{Long-Term Orbital Evolution}

JR1 is currently 2.7 AU away from Pluto and on a very similar orbit.
Because they are both in 3:2 resonant orbits with Neptune, both Pluto and JR1 are primarily perturbed by Neptune.
However, the combined Pluto system (Pluto and its five satellites) is by far the most massive 3:2 resonant KBO system.
We therefore sought to test what influence, if any, Pluto would have on JR1 during this future period of interactions.
We first selected 250 random state vectors from our unbiased cloud of orbits.
Using the same N-Body integrator and planetary initial states/masses as we used for fitting,
we integrated those state vectors first back and then forward 10 million years from the present,
saving the barycentric orbits for JR1 and Pluto along the way.
We then re-ran the same simulations, but for a control case with the mass of Pluto set to zero.

JR1 is primarily perturbed by Neptune, though it is not trapped in a Kozai resonance, as its perihelion circulates.
Pluto's argument of perihelion librates every 3.8 million years \citep[and citations therein]{1997plch.book..127M}, while JR1's argument circulates every 0.83 million years.
This combines with the nodal regression of the two bodies to be in conjunction every 2.4 million years ($\approx10^4$ orbits).
Because of Pluto's eccentric and inclined orbit, the Pluto-JR1 distance at one of these conjunctions is highly variable,
but can be as close 0.5 AU.
During the conjunctions, Pluto does perturb JR1, as can be seen in Figure \ref{fig:evo}.
But because the close approach distance is variable, Pluto's perturbation on JR1 is effectively chaotic.
While this could be described as Pluto quasi-satellite behavior \citep[and has:][]{2012MNRAS.427L..85D}, 
since JR1's orbit is still primarily controlled by Neptune,
a more precise description would be a series of periodic Pluto scattering events, one of which JR1 is currently experiencing.
Regardless of semantics, JR1 is certainly a case of Pluto exerting its gravitational influence over a fellow 3:2 resonant object.

\section{Summary}

(15810) 1994 JR$_1$ has a V-R of 0.76, making it a very red KBO.
Unique \textit{New Horizons} observations showed that JR1 has a high surface roughness of 37$\pm$5$^\circ$, indicating that it is potentially very cratered.
They also showed that the rotational period of JR1 is 5.47$\pm$0.33 hours, faster than most similar-sized KBOs,
and enabled a reduction of radial uncertainty of JR1's position from 10$^5$ to 10$^3$ km.
Neptune perturbations bring Pluto and JR1 close together every 2.4 million years, 
when Pluto can perturb JR1's orbit.
Future ground-based photometry of JR1 would be useful to better constrain the period and opposition surge,
and to allow preliminary estimates of JR1's shape and pole.
These proof of concept distant KBO observations demonstrate that the \textit{New Horizons} extended mission 
will indeed be capable of observing dozens of distant KBOs during its flight through the Kuiper Belt.

\acknowledgments

Special thanks to David Tholen for recovering JR1 after a 13-year gap, enabling all the observations in this paper,
and to Paul Helfenstein for the use of photometric modeling software.
This work was supported by NASA's New Horizons Project.
This work uses observations made with the NASA/ESA Hubble Space Telescope, and associated with program \#14092.
Support for this program was provided by NASA through a grant from the Space Telescope Science Institute, 
which is operated by the Association of Universities for Research in Astronomy, Inc., under NASA contract NAS 5-26555. 

\vspace{5mm}
\facilities{New Horizons(LORRI), HST(WFC3)}

\software{AstroPy, emcee, Matplotlib, NAIF SPICE, IDL}

\end{document}